\documentclass[twocolumn,showpacs,superscriptaddress,preprintnumbers,amsmath,amssymb]{revtex4}
\usepackage{graphicx}
\usepackage{dcolumn}
\usepackage{bm}
\usepackage{subfigure}
\usepackage{epsf}
\providecommand{\href}[2]{#2}

\def\simge{\mathrel{
     \rlap{\raise 0.511ex \hbox{$>$}}{\lower 0.511ex \hbox{$\sim$}}}}
\def\simle{\mathrel{
     \rlap{\raise 0.511ex \hbox{$<$}}{\lower 0.511ex \hbox{$\sim$}}}}

\begin{document}
\title{Does interferometry probe thermalization?}

\author{Cl\'ement Gombeaud}
\affiliation{Institut de Physique Th\'eorique, CEA/DSM/IPhT,
  CNRS/MPPU/URA2306\\ CEA Saclay, F-91191 Gif-sur-Yvette Cedex.}
\author{Tuomas Lappi}
\affiliation{Institut de Physique Th\'eorique, CEA/DSM/IPhT,
  CNRS/MPPU/URA2306\\ CEA Saclay, F-91191 Gif-sur-Yvette Cedex.}
\affiliation{Department of Physics
 P.O. Box 35, 40014 University of Jyv\"askyl\"a, Finland}
\author{Jean-Yves Ollitrault}
\affiliation{Institut de Physique Th\'eorique, CEA/DSM/IPhT,
  CNRS/MPPU/URA2306\\ CEA Saclay, F-91191 Gif-sur-Yvette Cedex.}
\date{\today}
\begin{abstract}
We carry out a systematic study of interferometry radii in
ultrarelativistic heavy-ion collisions within a two-dimensional
transport model. 
We compute the transverse radii $R_o$ and $R_s$ as a function of $p_t$
for various values of the Knudsen number, which measures the degree of
thermalization in the system. 
They converge to the hydrodynamical limit much more slowly (by a
factor $\simeq 3$) than elliptic flow. 
This solves most of the HBT puzzle for central collisions: 
$R_o/R_s$ is in the range $1.1-1.2$
for realistic values of the Knudsen number, much closer to
experimental data ($\simeq 1$) than the value $1.5$ from
hydrodynamical calculations.
The $p_t$ dependence of $R_o$ and $R_s$, which is usually said to
reflect collective flow, also has a very limited sensitivity to the
degree of thermalization. 
We then study the azimuthal oscillations of $R_o$, $R_s$, and $R_{os}$
for non central collisions. Their amplitudes depend little on the 
Knudsen number, and reflect the eccentricity of the overlap area
between the two nuclei. 
\end{abstract}
\pacs{25.75.Gz}
\maketitle
\section{Introduction}
Correlations of identical particles produced in ultrarelativistic
heavy-ion collisions have the unique capability to access directly the 
size of the fireball  \cite{Lisa:2005dd}. More precisely, they measure
the separation distribution of particles with a given momentum $\vec
p$ (regions   of homogeneity~\cite{Akkelin:1995gh}) after the last
interaction. These data, referred to as
HBT~\cite{Brown:1956zz}, thus impose severe constraints on model
calculations.  
In particular, blast-wave~\cite{Retiere:2003kf} and hydrodynamical
models~\cite{Kolb:2003dz,Hirano:2002ds,Zschiesche:2001dx,Socolowski:2004hw,Huovinen:2006jp},   
 which have been rather successful in reproducing transverse momentum spectra and
elliptic flows of identified particles up to $p_t\simeq 2$~GeV/c,
fail in reproducing HBT radii. 
More specifically, they generally overpredict the longitudinal size
$R_L$, as well as the ratio $R_o/R_s$, where $R_o$ and $R_s$
are the transverse radii parallel and orthogonal to the transverse
momentum, respectively. 
On the other hand, they correctly predict the decrease of radii with
$p_t$, which is often claimed to be a signature of collective flow.  
Viscous hydrodynamics gives smaller values of $R_o/R_s$ than
ideal hydrodynamics~\cite{Romatschke:2007jx,Pratt:2008qv}. 
Transport models also yield a smaller value, much closer to data,
typically around 1.2~\cite{Lin:2002gc,PhysRevLett.92.052301,Li:2006gp}. 

In this paper, we investigate systematically 
the sensitivity of HBT radii to the
degree of thermalization in the system. 
We explain the difference between predictions from hydrodynamics
(where local thermalization is assumed) and transport models (where
the system is generally not locally equilibrated). 
We consider a simple model, where the system consists of massless
particles undergoing $2\to 2$ elastic 
collisions~\cite{Gombeaud:2007ub}. The mean free path of a particle
between two collisions can be chosen arbitrarily by varying the cross
section. The limit of zero mean free path is the 
``hydrodynamic  limit'': the system is locally thermalized and its
expansion follows the laws of ideal hydrodynamics. The limit of
infinite mean free path corresponds to free-streaming particles: in
this case, HBT radii reflect the initial distribution of
particles. For finite values of the mean free path, the system is 
partially thermalized. 
In this paper, we study quantitatively how HBT observables vary
between these two extremes. 
A further simplification is that we consider only a two-dimensional
system living in the transverse plane: the present study is therefore
limited to the transverse radii $R_o$ and $R_s$ (and the cross term
$R_{os}$  for non-central collisions), and longitudinal expansion is
not taken into account. 

We do not mean to provide a realistic model of heavy-ion collisions. 
The fact that we consider only $2\to 2$ collisions implies, for sake 
of consistency, that the system is dilute (higher-order processes such
as $3\to 3$, are negligible), hence the equation of state is that of a 
perfect gas. This is the price to pay for a control handle on
thermalization. In the real world, the equation of state of QCD is not
that of a perfect gas: it has a very sharp  structure around $T\sim 
170$~MeV~\cite{Karsch:2003jg}, which is expected to influence   
observables, including HBT radii. By comparing with experimental data 
from RHIC, we expect to fail whenever the equation of state is
important. This will allow us to disentangle effects which can be
attributed to the equation of state, from those which are due to
thermalization and flow.

This article is organized as follows. In Sec.~\ref{s:model}, we
present our model and explain how HBT radii are obtained. 
Sec.~\ref{s:central} discusses the $p_t$ dependence of $R_o$ and $R_s$
and the value of $R_o/R_s$ for central collisions. 
Results from transport theory and ideal hydrodynamics are compared. 
Sec.~\ref{s:azhbt} discusses the azimuthal oscillations of $R_o$,
$R_s$ and $R_{os}$ for noncentral collisions. 
Our conclusions are summarized in Sec.~\ref{s:conclusions}.


\section{Model}
\label{s:model}

Nuclei colliding at RHIC are thin pancakes due to the strong Lorentz
contraction along the collision axis.
This large separation between the longitudinal and transverse scales
implies that longitudinal and transverse dynamics are to a large
extent decoupled. In this paper, we 
concentrate on the transverse expansion, which we model using a
2-dimensional relativistic Boltzmann equation~\cite{Gombeaud:2007ub}.  
We first describe the initial conditions of the evolution. 
We briefly recall how the Boltzmann equation is solved. 
We then define the Knudsen number, which measures how close the system
is to local thermal equilibrium. We finally define HBT radii.  

\subsection{Initial conditions}

The nucleus-nucleus collision creates particles. We assume for
simplicity that the spatial distribution of these particles is
initially a gaussian in the transverse plane:
\begin{equation}
\label{initialprofile}
n(x,y)=
\frac{N}{2\pi\sigma_x\sigma_y}e^{-\frac{x^2}{2\sigma_x^2}-\frac{y^2}{2\sigma_y^2}},
\end{equation}
where $N$ is the total number of particles, and $\sigma_x$ and
$\sigma_y$ are the rms widths of the distributions in the $x$ and $y$
directions. The $x$ axis denotes the direction of impact parameter, or
reaction plane. 

As for the initial momentum distribution, two different scenarios have
 been implemented and compared. 
The first scenario is the same as in~\cite{Gombeaud:2007ub}. 
In order to compare transport theory and hydrodynamics, we take the
 same initial conditions:
The momentum  distribution is locally thermal, and the temperature is
 related to the density according to the equation of state of a
 2-dimensional massless, ideal gas: 
 $T\propto n^{1/2}$. Since our calculation is purely classical, we
 assume Maxwell-Boltzmann statistics for sake of consistency:
\begin{equation}
\label{thermal}
\frac{dN}{d^2 pd^2 x}\propto \exp\left(-\frac{p}{T(x,y)}\right)
\end{equation}
where $T(x,y)$ is the local temperature, given by:
\begin{equation}
\label{Tprofile}
T(x,y)=T_0\exp\left(-\frac{x^2}{4\sigma_x^2}-\frac{y^2}{4\sigma_y^2}\right). 
\end{equation}

The second set of initial conditions are taken from
the Color glass condensate (CGC) calculations~\cite{Krasnitz:2001qu,Lappi:2003bi},
where the initial gluon spectrum is calculated by solving the classical 
Yang-Mills (CYM) equations with the initial conditions given by the MV 
model~\cite{McLerran:1994ni}. The result of the numerical computation
can be parameterized~\cite{Krasnitz:2003jw} as
\begin{equation}
\label{cgc}
\frac{dN}{d^2 p_td^2 x}=\left\{
\begin{array}{ll}
a_1[e^{\frac{p_t}{b \Lambda_s}}-1]^{-1} & (p_t/\Lambda_s)<1.5\\
a_2\log(4\pi p_t/\Lambda_s)(p_t/\Lambda_s)^{-4} & (p_t/\Lambda_s)>1.5
\end{array}
\right.
\end{equation}
with $a_1=0.137$, $a_2=0.0087$ and $b=0.465$.
The color charge density parameter $\Lambda_s$ (proportional to the saturation scale 
$Q_s$~\cite{Lappi:2007ku}) plays the role of the temperature as the only 
transverse momentum scale in the system.
The parameterization (\ref{cgc}) was fit to a calculation for a nucleus of an
infinite size on the transverse plane, but we generalize it by letting
$\Lambda_s(x,y)$ have the same Gaussian dependence on the transverse coordinate
as the temperature in Eq.~(\ref{Tprofile}), with an absolute value adjusted
to give the same value for $\langle p_t\rangle$.

Our 2-dimensional kinetic theory approach does not contain longitudinal 
expansion, and therefore cannot address questions related to isotropization
of the particle distribution. The CGC initial conditions naturally lead
to a very anisotropic initial condition where, after $\tau\sim 1/Q_s$,
$\langle p_z \rangle \ll \langle p_t\rangle$, whereas conventional 3
dimensional hydrodynamics assumes isotropy in the local rest frame. 
In the 2-dimensional approach, $p_z=0$ for all particles, and the
energy per particle $\langle p_t\rangle$ is 
constant throughout the evolution; we adjust it to $\langle
p_t\rangle=420$~MeV,  corresponding roughly to the value for pions at
the top RHIC energy~\cite{Adler:2003cb}.  
This fixes the value of $T_0$ for the thermal initial conditions 
(\ref{Tprofile}) 
($\langle p_t\rangle=\frac{4}{3}T_0$) 
and the value of $\Lambda_s$ in Eq.~(\ref{cgc}) for the CGC initial
conditions. 
In practice, this implies an unrealistically small value of the
saturation scale $Q_s$. 
Conventional estimates of $Q_s$ are larger, 
but significant longitudinal cooling is required in order to match
observed $p_t$ spectra.  
In our calculation, the conservation of $\langle p_t\rangle$ 
makes our initial $\langle p_t\rangle$ smaller than most estimates. 
However, our emphasis 
in this paper is on the influence of thermalization on the \emph{transverse} HBT radii
and the \emph{transverse} momentum dependence. For this purpose the two initial conditions 
that we use represent the opposite ends of the range of physically reasonable 
$p_t$ spectra: from fully thermalized to one with a perturbative power law 
behavior at large $p_t$ that one would expect to observe in the absence of any final 
state interactions.

\subsection{Expansion: Knudsen number}

The results presented in this paper use the algorithm described
in~\cite{Gombeaud:2007ub} to solve the two-dimensional relativistic
Boltzmann equation. 
The Monte-Carlo algorithm follows the trajectory of every particle
throughout the expansion of the system, until they cease to interact. 
Particles interact through $2\to 2$ elastic collisions. 
The cross section is assumed isotropic in the center-of-mass frame for
simplicity. 

The only remaining parameters in the simulation are the total number
of particles, $N$, and the elastic cross section,
$\sigma$. (Since we are working in two dimensions, $\sigma$ has
  the dimension of a length.)
The initial average particle density per unit surface is:
\begin{equation}
\bar n=\frac{N}{4\pi\sigma_x\sigma_y}.
\end{equation}
It is worth emphasizing that $N$ is usually much larger in the
Monte-Carlo simulation than in an actual heavy-ion 
collision (``parton subdivision'' technique). This can be understood 
in the following way. 
The physical length scale that should be independent of 
$N$ is the mean free path $\lambda$. Its precise value depends on the 
initial velocity and position of the particle, but the order
of magnitude is generally $\lambda=1/\sigma\bar n$. 
$\lambda$  must be compared to another length scale, 
the average interparticle distance $d=\bar{n}^{-1/2}$. 
Let us define the dilution parameter $D$ as 
\begin{equation}
\label{dilution}
D=\frac{d}{\lambda}=\sigma \bar{n}^{1/2} = \frac{1}{\lambda \bar{n}^{1/2}}.
\end{equation}
Our description of the system in terms of elastic $2 \to 2$ collisions is
consistent only in the limit when $D$ is small and the contribution of
many-body collisions is suppressed. 
This is a requirement of the Boltzmann equation~\cite{Gombeaud:2007ub}.
To achieve this one must take the limit
of large $N$ and small $\sigma$ keeping $\sigma N$ fixed. This ensures
that our results are extrapolations to the limit $N\to \infty$ at
fixed $\lambda$. 
Because in this limit $\sigma$ approaches zero the
interactions between the particles become truly pointlike, 
and problems with causality and Lorentz-invariance are avoided.
For this reason, all the results
presented in this paper are obtained by doing two simulations with the
same value of $\lambda$ and different values of $D$; the results are
then extrapolated linearly to $D=0$. 

The standard dimensionless parameter to characterize the degree of
thermalization is the Knudsen number $K$, defined as the ratio of the  
mean free path to the characteristic size of the system $R$. 
We define $R$ as in~\cite{Bhalerao:2005mm}:
\begin{equation}
R=\left(\frac{1}{\sigma_x^2}+\frac{1}{\sigma_y^2}\right)^{-1/2}.
\end{equation}
The Knudsen number $K$ is then defined as 
\begin{equation}
\label{defKn}
K\equiv \frac{\lambda}{R}=\frac{1}{\sigma n R}.
\end{equation}
The inverse of the Knudsen number is proportional to the average
number of collisions per particle, $n_{\rm coll}$. The product $n_{\rm
  coll} K$ remains very close to 1.6, for all values of
$K$~\cite{Gombeaud:2007ub}. 
Hydrodynamics is the limit $K\to 0$, while $K\to +\infty$ correspond to free 
streaming particles. 
A fit to the centrality dependence of elliptic
flow~\cite{Drescher:2007cd} suggests that $K\simeq 0.3$ for central
Au-Au collisions at RHIC. 

We choose to keep the scattering cross section $\sigma$ constant as a
function of time for sake of simplicity, as for instance in the AMPT
transport model~\cite{Lin:2004en}. Other transport calculations have
been carried out~\cite{Huovinen:2008te} where the viscosity to entropy
ratio $\eta/s$ is kept constant, so that $\sigma$ depends on
temperature or time, typically like $t^{2/3}$. 
As we recall below, HBT radii give a measure of the system when the
last scattering occurs, that is, much later than other observables such
as elliptic flow. Therefore, results might differ significantly with a
time-dependent cross section.

\subsection{HBT radii}
\label{HBTparam}

For a particle with momentum ${\bf p}_t$, we denote by $(t,x,y)$ the
space-time  point where the last collision occurs. The ``out'' and
``side'' coordinates are then defined as the projections parallel and 
orthogonal to the particle momentum:
\begin{eqnarray}
\label{defxos}
x_o&=&{\bf x}\cdot{\bf v}-v t=x\cos\phi+y\sin\phi- v t\cr
x_s&=&{\bf x}\times {\bf v}=x\sin\phi-y\cos\phi,
\end{eqnarray}
where ${\bf v}\equiv {\bf p}_t/p_t$ is the particle velocity 
($v=1$), and $\phi$ its azimuthal angle:
${\bf p}_t=(p_t\cos\phi,p_t\sin\phi)$. 
Both $x_o$ and $x_s$ are invariant under a translation along the
trajectory after the last scattering: $(t,\vec x)\to (t+\tau,\vec
x+\vec v\tau)$. In particular, they are invariant through a scattering
at zero angle.  
HBT radii are obtained by averaging over many particles with the
same momentum:
\begin{eqnarray}
\label{HBTdefine}
R_o^2 & = & \langle x_o^2\rangle-\langle x_o\rangle^2\cr
R_s^2 & = & \langle x_s^2\rangle-\langle x_s\rangle^2\cr
R_{os} & =& \langle x_o x_s\rangle - \langle x_o\rangle \langle x_s\rangle.
\end{eqnarray}
Radii defined in this way coincide with those obtained from the 
curvature of the correlation function at zero relative momentum, in the
absence of final-state interactions~\cite{Lin:2002gc}. 
Experimentally, radii are usually obtained from gaussian fits to the 
correlation function. This procedure gives different radii if the
source is not gaussian~\cite{Lisa:2005dd}. In our case, we have
checked explicitly that sources are close to gaussian, but we have not
investigated systematically effects of non gaussianities.  
Strictly speaking, averages in Eq.~(\ref{HBTdefine}) are for a given
momentum. In practice, our results are obtained 
by taking bins of width $10$~MeV/c in $p_t$, and averaging over the
particles in the bin. We have checked that  
results do not vary significantly with a smaller bin size.

The radii defined by Eq.~(\ref{HBTdefine}) are generally functions of
$p_t$ and $\phi$. 
In Sec.~\ref{s:central}, we study central collisions,
with $\sigma_x=\sigma_y$. Symmetry of the system with respect to the
direction of ${\bf
  p}_t$ then implies $R_{os}=0$. Rotational symmetry implies that
$R_o$ and $R_s$ are independent of $\phi$. The more general case when 
$\sigma_x$ and $\sigma_y$ differ is studied in Sec.~\ref{s:azhbt}. 


\section{Central collisions}
\label{s:central}

In this section, we discuss how the $p_t$ dependence of $R_o$ evolves
with the Knudsen number for a central collision. 
We then discuss the ratio $R_o/R_s$. Finally, we compare with
experimental data. 
In order to mimic a central Au-Au collision at RHIC, we use the
initial density profile Eq.~(\ref{initialprofile}) with
$\sigma_x=\sigma_y=3$~fm. This value corresponds to the rms width of
the initial density profile in an optical Glauber
calculation~\cite{Miller:2007ri}.

\begin{figure}
\includegraphics*[width=\linewidth]{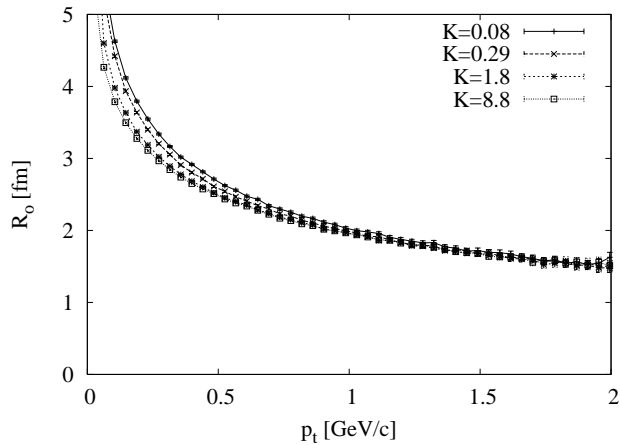}
\caption{HBT radius $R_o$ versus  transverse momentum $p_t$ of
  particles in the transport calculation. The curves are labeled by
  the value of the Knudsen number $K$.} 
\label{fig:fig1}
\end{figure}
The decrease of HBT radii with the transverse momentum $p_t$,
typically like $p_t^{-1/2}$, is
often~\cite{Lisa:2005dd,Makhlin:1987gm,Ollitrault:2002zy} 
presented as a signature of collective flow. 
Collective flow is associated with the hydrodynamic limit, i.e., the
limit of small $K$. Fig.~\ref{fig:fig1} displays $R_o$ versus $p_t$
for thermal initial conditions, Eq.~(\ref{thermal}), and several 
values of the Knudsen number $K$.  
Generally, $R_o$ increases as $K$ decreases. However, this is a small
effect. 
The decrease of $R_o$ with $p_t$ is more pronounced in the
hydrodynamic limit (small $K$) but is also seen for free streaming 
particles (large $K$). 
For large $K$, HBT radii reflect the initial momentum distribution:
both with thermal initial conditions
conditions, Eq.~(\ref{thermal}), and with CGC initial conditions,
Eq.~(\ref{cgc}), particles with higher $p_t$ are more likely
to be produced in dense regions, i.e., near the center of the fireball
$x=y=0$. 
For $p_t\gg \langle p_t\rangle$, Eqs.~(\ref{thermal}) and
(\ref{Tprofile}) yield $R_o(p_t)\simeq \sigma_x\sqrt{1.5\langle
  p_t\rangle/p_t}$ for the initial distribution, while 
Eq.~(\ref{cgc}) gives  $R_o(p_t)\simeq \sigma_x/\sqrt{2}$. 
In practice, after collisions have occurred, both sets of initial
conditions yield similar radii, as we shall see explicitly
later. 

\begin{figure}
\includegraphics*[width=\linewidth]{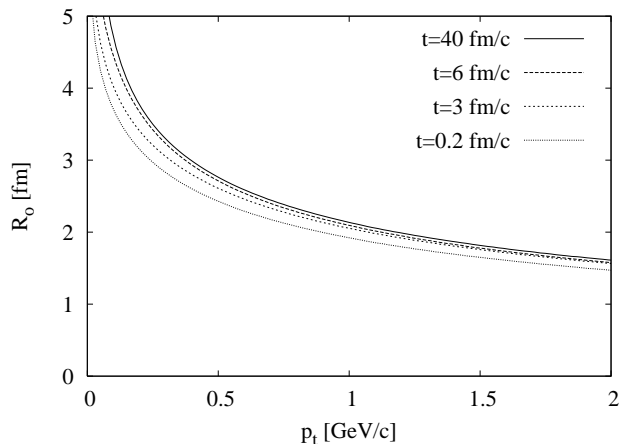}
\caption{HBT radius $R_o$ versus transverse momentum $p_t$ of
  particles in ideal hydrodynamics, at a given time $t$. The curves
  are labeled by the value of $t$.} 
\label{fig:fig2}
\end{figure}
Decreasing the Knudsen number $K$ amounts to increasing the number of
collisions, hence the ``freeze-out'' time when the last collision
occurs. To show this explicitly, Fig.~\ref{fig:fig2} displays 
$R_o(p_t)$ in ideal hydrodynamics, assuming sudden freeze-out at time $t$. 
The equations of hydrodynamics are solved using a first-order Godunov
scheme~\cite{Ollitrault:1992bk}. 
For sake of consistency with the
transport calculation, the equation of state of the fluid is that of a
two-dimensional ideal gas, and there is no longitudinal
expansion~\cite{Gombeaud:2007ub}.
Hydrodynamics at $t=0.2$~fm/c gives the same radii as the transport
calculation for 
large $K$ in Fig.~\ref{fig:fig1}, because we have chosen the same
initial conditions for both calculations. 
As time evolves, $R_o$ increases; the increase is more pronounced
and occurs later at low $p_t$. At a given $p_t$, the value of $R_o$
converges as $t$ increases. This is by no means a trivial result: 
the location of the last interaction, $\langle x_o\rangle$ in
Eq.~(\ref{HBTdefine}), increases linearly with $t$. Only the
dispersion $R_o$ of this location converges. 
Hydrodynamics at large $t$ is almost identical to transport at small
$K$ (the relative difference is less than 5\%). 
This is also a non-trivial result, although it is implicit in all
hydrodynamical studies of HBT observables~\cite{Kolb:2003dz}:
HBT observables are defined at the last scattering, when the system is
no longer in local equilibrium, and hydrodynamics is not valid.

\begin{figure}
\resizebox{3.5in}{!}{
\includegraphics*[width=\linewidth]{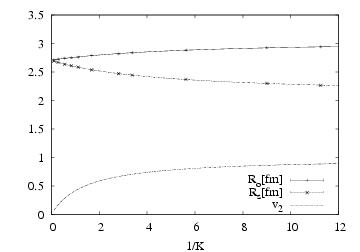}}
\caption{$R_o$ and $R_s$, averaged over the interval 
  $0.25<p_t<0.75$~GeV/c, versus $1/K$, which scales like the number of
  collisions per particle. The lines are 3-parameter fits with
  Eq.~(\ref{fitformula}). The dotted curve shows, for sake of
  illustration, the variation of elliptic flow in a non-central
  collision, scaled by the hydrodynamical limit
  (from~\cite{Gombeaud:2007ub}).}
\label{fig:fig3}
\end{figure}
Hydrodynamical calculations usually yield a value of $R_o/R_s$ which
is much too large, of the order of 1.5, while RHIC data are compatible
with 1. 
While $R_o$ increases with time in hydrodynamics, $R_s$
decreases. Initially, $R_o(p_t)=R_s(p_t)$ by symmetry. 
In the transport calculation, the same behavior is observed as the
number of collisions per particle $1/K$ increases, as shown in
Fig.~\ref{fig:fig3}.  
However, this increase is quite slow. The same curve shows, for sake
of illustration, the increase of the elliptic flow $v_2$ with $1/K$ for a
noncentral collision~\cite{Gombeaud:2007ub}. \emph{Elliptic 
flow converges to the ``hydrodynamic limit'' much faster than HBT
radii.}  In order to put this statement on a quantitative 
basis, we fit our numerical results for $R_o(K)$ with the following
formula~\cite{Gombeaud:2007ub}:  
\begin{equation}
\label{fitformula}
R_o(K)=R_o^{f.s.}+\frac{R_o^{\rm hydro}-R_o^{f.s.}}{1+K/K_0}.
\end{equation}
The fit parameters are the free-streaming ($K\to\infty$) limit 
$R_o^{f.s.}$, the hydrodynamic ($K\to 0$) limit $R_o^{\rm hydro}$, and
$K_0$, the value of the Knudsen number for which $R_o(K)$ is half-way
between free-streaming and hydro. A similar formula can be used for
$R_s(K)$. 
It fits our numerical results perfectly (see Fig.~\ref{fig:fig3}). 
The value of $K_0$ is $0.167\pm 0.007$ for $R_o$ and $0.215\pm 0.006$
for $R_s$, while it is $0.7$ for $v_2$~\cite{Gombeaud:2007ub}:
convergence toward the hydrodynamic limit requires 3-4 times more
collisions for HBT radii than for elliptic flow. 
The other fit parameters are $R_o^{\rm hydro}=3.071\pm 0.006$~fm and 
$R_s^{\rm hydro}=2.088\pm 0.006$~fm: we recover the HBT puzzle
$R_o/R_s\simeq 1.5$ in the hydrodynamical limit. 

\begin{figure}
\resizebox{3.5in}{!}{
\includegraphics*[width=\linewidth]{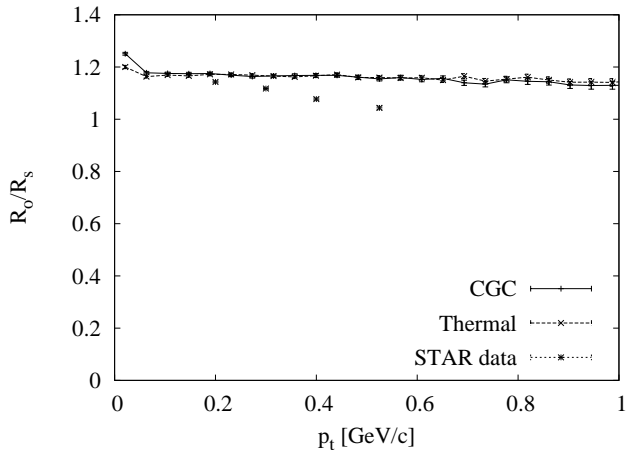}}
\caption{$R_o/R_s$ versus $p_t$ for $K=0.3$. Dashed line: thermal initial
  conditions, Eq.~(\ref{thermal}); full line: CGC
  initial conditions,  Eq.~(\ref{cgc}). Error bars on
  the transport calculations are statistical only. Stars: data
  from STAR~\cite{Adams:2004yc}.}
\label{fig:fig4}
\end{figure}
We now discuss the value of $R_o/R_s$ at RHIC. 
The centrality dependence of $v_2$ suggests that
$K\simeq 0.3$ in central Au-Au collisions~\cite{Drescher:2007cd}. 
For this value, $v_2$ is already 70\% of the hydrodynamic limit. On
the other hand, $R_o/R_s\simeq 1.16$, which is significantly below the
hydrodynamic limit of $1.5$. 
A similar value ($R_o/R_s\simeq 1.2$) was found with the AMPT transport
code~\cite{Lin:2002gc}.  
A more detailed comparison is shown in Fig.~\ref{fig:fig4}, which
displays $R_o/R_s$ versus $p_t$ for the two  
sets of initial conditions, Eqs.~(\ref{thermal}) and (\ref{cgc}). 
$K$ has been fixed to the value which is favored by 
$v_2$ data~\cite{Drescher:2007cd}, i.e., $K=0.3$ for central
collisions. For both sets of initial conditions, $R_o/R_s$ is
essentially independent of $p_t$, while data show a slight decrease. 
In this respect, our results differ from the covariant MPC model
(which is in principle equivalent to ours, with the longitudinal
expansion taken into account), where $R_o/R_s$ is found smaller than 1
at large $p_t$~\cite{PhysRevLett.92.052301}.  

Our value of $R_o/R_s$ for central collisions is in much better
agreement with experimental data than models based on ideal
hydrodynamics, which give a value around $1.5$. 
It has been recently argued that ideal hydrodynamics with an early
freeze-out~\cite{Broniowski:2008vp} also explains
the HBT puzzle. Our results also show that increasing the
Knudsen number amounts to decreasing the freeze-out time
in ideal hydrodynamics. 
Viscous corrections to ideal hydrodynamics, which incorporate 
deviations to local equilibrium to first order in the Knudsen number 
$K$, also lead to a reduced $R_o/R_s$ (together with a reduced
longitudinal radius $R_L$, also closer to data). However, the value of
$\eta/s$ required to match~\cite{Romatschke:2007jx} the data is much
larger than that inferred from the study of elliptic
flow~\cite{Romatschke:2007mq}, so that viscous corrections explain
only a small part of the HBT puzzle~\cite{Pratt:2008qv}. 
However, viscous hydrodynamics itself breaks down at freeze-out, and
may not be a reliable tool for estimating HBT radii. Our results
suggest that deviations from equilibrium have larger effects on HBT
radii than inferred from viscous hydrodynamics. We find that partial
thermalization, which has been shown to explain the centrality
dependence of $v_2$, also solves most of the HBT puzzle for $R_o/R_s$.

\begin{figure}[ht]
\includegraphics*[width=\linewidth]{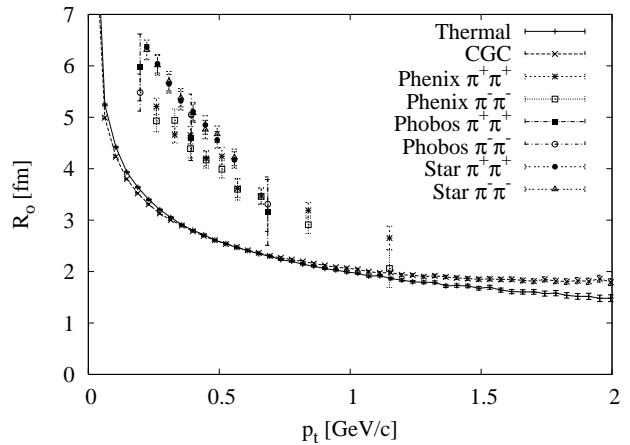}
\caption{HBT radius $R_o$ versus transverse momentum $p_t$. 
Lines: our results for $K=0.3$ and different initial
conditions: thermal, Eq.~(\ref{thermal}), and CGC, Eq.~(\ref{cgc}). 
Symbols: experimental data from
  STAR~\cite{Adams:2004yc}, PHOBOS~\cite{Back:2004ug} and
  PHENIX~\cite{Adler:2004rq}.}
\label{fig:fig5}
\end{figure}

While our transport calculation gives a plausible explanation for the
small $R_o/R_s$, it completely misses the absolute value of HBT
radii. This is shown in Fig.~\ref{fig:fig5}, which displays a
comparison between $R_o(p_t)$ from our transport calculation with data
from STAR~\cite{Adams:2004yc}, PHOBOS~\cite{Back:2004ug} and
PHENIX~\cite{Adler:2004rq}. 
Experimental values are much larger. 
This is  due to the equation of
state~\cite{Pratt:2008sz,Zschiesche:2001dx}, which is that of an 
ideal gas in our calculation. Our HBT volume $R_oR_s$ is 
essentially independent of the number of collisions $1/K$. It has been
argued that this is a general result for an ideal gas, due to entropy
conservation~\cite{Akkelin:2004he}. 
The equation of state of QCD, on the other hand, has a sharp 
structure around $T_c\sim 170$~MeV: as the temperature decreases, the
entropy density drops by an order of magnitude in a narrow interval
around $T_c$. In a heavy-ion collision, the volume increases by a
large factor with essentially no change in the temperature. This
explains why HBT volumes increase at the transition. 
A complete study must also take into account the longitudinal
expansion, and its effect on the longitudinal radius $R_L$.

\section{Azimuthally sensitive HBT}
\label{s:azhbt}

For a non-central collision, the interaction region is elliptic, and
HBT radii depend on $\phi$. 
Azimuthally-sensitive interferometry has been investigated
theoretically within hydrodynamical
models~\cite{Heinz:2002sq,Tomasik:2005ny,Frodermann:2007ab,Csanad:2008af,Kisiel:2008ws}
and transport models~\cite{Humanic:2005ye}.  
We first briefly recall why and how radii depend on $\phi$.  
We then study how the various radii depend on the Knudsen number.
Finally, we introduce dimensionless ratios of oscillation amplitudes,
which do not seem to have not been studied previously, and we compare
our results with experimental data~\cite{Adams:2004yc}. 

\begin{figure}[ht]
\includegraphics*[width=0.4\linewidth]{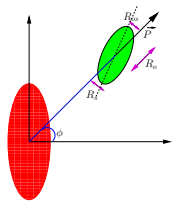}
\caption{Illustration of $\phi$ dependent HBT radii.}
\label{fig:fig6}
\end{figure}
 Fig.~\ref{fig:fig6} illustrates the $\phi$ dependence of transverse
radii. The initial distribution of matter is elongated along the $y$
axis, so that $R_o$ has a maximum at $\phi=\pi/2$, while $R_s$
has a maximum at $\phi=0$. 
Finally, $R_{os}$ differs from zero when the principal axes of 
the region of homogeneity are tilted relative to the direction of
momentum. 

Before we present our results, let us briefly explain how the 
$\phi$ dependence of the radii is evaluated in the Monte-Carlo
solution of the Boltzmann equation. The radii (\ref{HBTdefine})
involve average values, such as $\langle x_o\rangle$, which depend on
$\phi$. Such averages can be computed by binning in $\phi$, and
computing the average in each bin:
\begin{equation}
\label{phiaverage}
\langle x_o\rangle=\frac{\sum_{i=1}^N (x_o)_i}{N}, 
\end{equation}
where $N$ is the number of particles in the bin, which depends on
$\phi$ due to elliptic flow. Since the $\phi$ dependence is smooth,
more accurate results are obtained by fitting both the numerator and the 
denominator of Eq.~(\ref{phiaverage}) by Fourier series, using the
symmetries $\phi\to -\phi$ and $\phi\to \phi+\pi$ to restrain the
number of terms~\cite{PhysRevC.66.044903}. 
Since the Fourier expansion converges rapidly, we keep terms only up
to order $\cos 4\phi$ and $\sin 4\phi$.

\begin{figure}[ht]
\includegraphics*[width=\linewidth]{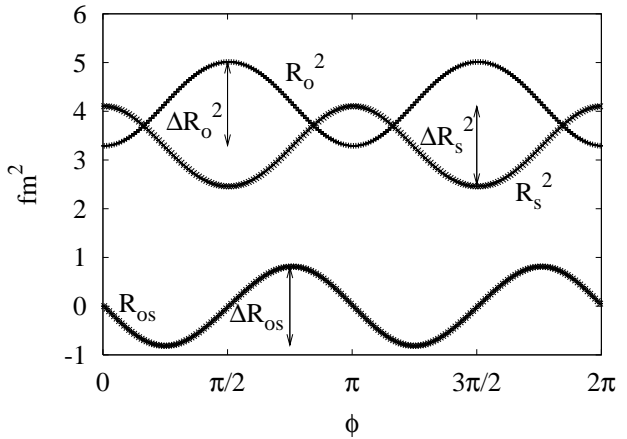}
\caption{Azimuthal dependence of HBT Radii. Thermal initial conditions,
with $\sigma_x=1.95$~fm, $\sigma_y=2.6$~fm, corresponding roughly to a
Au-Au collision at RHIC with impact parameter $b=7$~fm. 
The $p_t$ interval is the same as in Fig.~\ref{fig:fig3}. 
The Knudsen number is $K=0.4$.}
\label{fig:fig7}
\end{figure}
Fig.~\ref{fig:fig7} displays the $\phi$ dependence of $R_o^2$,
$R_s^2$, and $R_{os}$. 
The variation of $R_o^2$ and $R_s^2$ is clearly
dominated by a $\cos 2\phi$ term, while the variation of $R_{os}$ goes
like $\sin 2\phi$. The mean value of $R_o^2$ is slightly higher than
the mean value of $R_s^2$, which is not surprising since final-state
interactions increase $R_o$ and decrease $R_s$. At $\phi=0$, however,
$R_o<R_s$, reflecting the initial eccentricity of the system. 

\begin{figure}[!htb]
\resizebox{3.5in}{!}{
\includegraphics*[width=\linewidth]{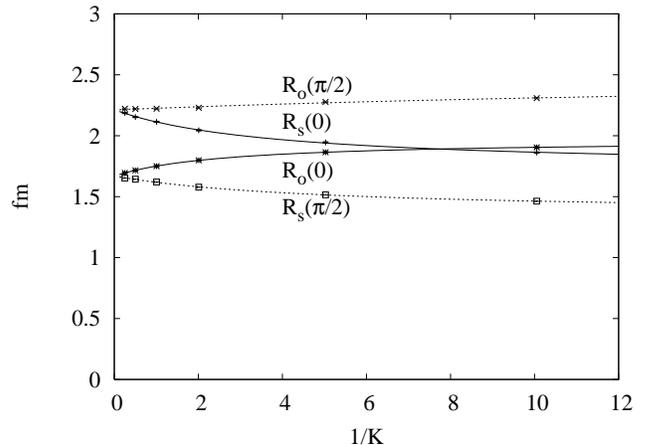}}
\caption{In-plane ($\phi=0$) and out-of-plane ($\phi=\pi/2$ radii
  versus $K^{-1}$ for thermal initial conditions. 
Initial conditions and $p_t$ interval as in
  Fig.~\ref{fig:fig7}.  
Lines are 3-parameter fits using Eq.~(\ref{fitformula}). 
}  
\label{fig:fig8}
\end{figure}
Fig.~\ref{fig:fig8} displays radii in the reaction plane ($\phi=0$)
and out of the reaction plane ($\phi=\pi/2$) versus $K$. 
$R_o$ increases and $R_s$ decreases as the number of collisions $1/K$
increases, as already observed for central collisions
(Fig.~\ref{fig:fig3}). Upon closer scrutiny, Fig.~\ref{fig:fig8}
reveals that the slope of the curves differ. This is reflected by the
value of the 
parameter fit $K_0$ in Eq.~(\ref{fitformula}). $K_0$ is largest for
$R_o(0)$ ($K_0=0.38\pm 0.01$), smallest for $R_o(\pi/2)$
($K_0=0.04\pm 0.05$), and intermediate for $R_s(0)$ and $R_s(\pi/2)$
($K_0=0.26\pm 0.03$ and $0.20\pm 0.02$, respectively). Our 
interpretation is that thermalization is faster in plane than out of
plane, which is natural since collective flow is preferentially in plane.  

We now study quantitatively how oscillation amplitudes vary with
$K$. There are three such amplitudes, as illustrated in Fig.~\ref{fig:fig7}:
\begin{eqnarray}
\label{oscamplitudes}
\Delta R_o^2 &=& R_o^2(\pi/2)-R_o^2(0)\cr
\Delta R_s^2 &=& R_s^2(0)-R_s^2(\pi/2)\cr
\Delta R_{os} &=& R_{os}(3\pi/4)-R_{os}(\pi/4).
\end{eqnarray}
In Fig.~\ref{fig:fig7}, all three amplitudes are clearly comparable. 
If $K\gg 1$, particles escape freely after they have been produced. Setting
$t=0$ in Eq.~(\ref{defxos}) and using the fact that the initial
distribution is centered at $x=y=0$ and has $y\to -y$ symmetry, one
easily shows that all three amplitudes are equal to $\langle
y^2-x^2\rangle$.
The results are integrated over the $p_t$ range $0.25<p_t<0.75$~GeV/c,
but our results depend weakly on $p_t$. 
In particular, we do not see the inversion of oscillations 
at large $p_t$ reported in earlier hydrodynamical 
calculations~\cite{Heinz:2002sq,Frodermann:2007ab}. This inversion was
not observed in more recent calculations~\cite{Kisiel:2008ws}.

\begin{figure}[!htb]
\resizebox{3.5in}{!}{
\includegraphics*[width=\linewidth]{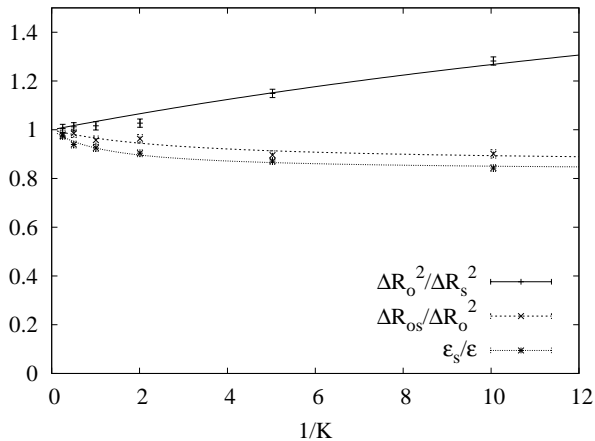}}
\caption{Ratios of oscillation amplitudes versus $K^{-1}$ for thermal
  initial conditions. $R_o^2$, $R_s^2$ and $R_{os}$ are integrated
  over the $p_t$ interval $0.5<p_t<0.75$~GeV/c.
Lines are drawn to guide the eye. }  
\label{fig:fig9}
\end{figure}
Oscillation amplitudes scale like the eccentricity of the overlap area
between the two nuclei, which depends on centrality and is not known
directly. This dependence can be avoided by considering {\it ratios\/}
of oscillation amplitudes. Out of $3$ amplitudes, one may construct
$2$ ratios, $\Delta R_{os}/\Delta R_o^2$ and $\Delta R_o^2/\Delta
R_s^2$. These ratios can be extracted directly from experimental data, 
and are equal to unity in the free-streaming limit (large $K$). 
They are plotted in Fig.~\ref{fig:fig9} versus $1/K$. 
Final-state interactions increase the oscillations of $R_o$ relative
to $R_s$, much in the same way as they increase $R_o$ relative to
$R_s$ for central collisions.
The opposite behavior was found in hydro \cite{Frodermann:2007ab} and
blast-wave~\cite{Retiere:2003kf} calculations, and we do not
understand the origin of this discrepancy. 

For realistic values of $K$, both ratios deviate little from unity. 
It is also interesting to compare the eccentricity seen in HBT radii,
for instance in $R_s$:
\begin{equation}
\label{HBTeps}
\epsilon_s\equiv \frac{R_{s}^2(0)-R_{s}^2(\pi/2)}{R_{s}^2(0)+R_{s}^2(\pi/2)}
\end{equation}
with the initial eccentricity  
\begin{equation}
\label{defepsilon}
\epsilon=\frac{\sigma_y^2-\sigma_x^2}{\sigma_y^2+\sigma_x^2}.
\end{equation}
In the limit $K\to\infty$,  $\epsilon_s$ and $\epsilon$ are strictly
equal for both sets of initial conditions. 
The ratio  $\epsilon_s/\epsilon$ is plotted in Fig.~\ref{fig:fig9}. 
It also remains close to unity.
We conclude that none of the observables we can construct from 
oscillation amplitudes is an interesting probe of thermalization. 
For realistic values of $K$, the region of homogeneity essentially
retains the shape of the initial distribution. 

\begin{table}
\begin{tabular}{||c||c||c|c||}
\hline
  & {STAR data} & \multicolumn{2}{c||}{our results} \\
\cline{3-4}
   & & $K=0.32$ & $K=0.51$  \\
\hline
$\Delta R_o^2/\Delta R_s^2$ & $1.45 \pm 0.61 $ &$ 1.08 \pm 0.02$  &$ 1.05 \pm 0.02$   \\  
\hline
$\Delta R_{os}/\Delta R_o^2$ &$ 0.68 \pm 0.42$ &$ 0.97 \pm 0.03$ &$ 0.99 \pm 0.03$ \\
\hline
$\epsilon_s$ & $0.080 \pm 0.026$ &$ 0.205 \pm 0.003$ &$ 0.213 \pm 0.005$ \\

\hline
\end{tabular}
\begin{tabular}{||c||c||c|c||}
\hline
  & {STAR data} & \multicolumn{2}{c||}{our results} \\
\cline{3-4}
   & & $K=0.31$ & $K=0.49$  \\
\hline
$\Delta R_o^2/\Delta R_s^2$ & $1.09 \pm 0.46$  &$ 1.14 \pm 0.02$  & $1.06 \pm0.02$\\  
\hline
$\Delta R_{os}/\Delta R_o^2$ &$ 0.65 \pm 0.31$ &$ 0.90 \pm 0.04$ &$ 0.92 \pm 0.04$ \\
\hline
$\epsilon_s$ &$ 0.086 \pm 0.017$ &$ 0.172 \pm 0.005$ &$ 0.174 \pm 0.005$ \\
\hline
\end{tabular}
\caption{Comparison between results from STAR~\cite{Adams:2004yc} and our
  calculations. Top: centrality interval 20-30\% and $k_t\in 
  [0.15,0.25]$ GeV/c. Bottom: centrality interval
  10-20\% and $k_t\in [0.35,0.45]$ GeV/c. } 
\label{tab:table}
\end{table} 
Although our model calculation is too crude to reproduce the magnitude
of HBT radii, we expect that the above ratios are less model
dependent;  in particular, they all go to 1 in the absence of 
final-state interactions. A comparison with existing data is therefore
instructive. 
Table~\ref{tab:table}  displays comparisons
between our results and experimental data from
STAR~\cite{Adams:2004yc}. 
The correspondence between centrality and eccentricity was taken
from~\cite{Kolb:2001qz}. For each set of data, the two values of the
Knudsen number span the range inferred from the centrality dependence
of $v_2$~\cite{Drescher:2007cd}. 
Note that the $p_t$ ranges differ for the two centrality
intervals. This is the reason why our results are also slightly
different, although the values of $K$ are essentially the same. 
Our results for $\Delta R_o^2/\Delta R_s^2$ and $\Delta
R_{os}/\Delta R_o^2$ are compatible with experimental data, but the
latter have large error bars. On the other hand, the experimental
value of $\epsilon_s$ is smaller by a factor 2 than our
value. In our calculations, $\epsilon_s$ remains very close to the
initial eccentricity (see Fig.~\ref{fig:fig9}). Experimentally,
however, the initial eccentricity seems to be washed out by the
expansion. 
This is a spectacular effect, whose importance doesn't seem to
  have been fully appreciated so far.
Hydrodynamical calculations have been reported~\cite{Kisiel:2008ws}
which are in fair agreement with the measured value of
$\epsilon_s$. These calculations use a soft equation of state: 
it is likely that the soft equation of state of QCD is responsible for the
reduced eccentricity seen in data. 

\section{Conclusions}
\label{s:conclusions}

We have carried out a systematic study of how HBT observables evolve
with the degree of thermalization in the system, characterized by the
Knudsen number $K$. The number of collisions per particle scales like
$1/K$, and local equilibrium corresponds to the limit $K\to 0$. 
Our results show that HBT observables depend very weakly on $K$:
\begin{itemize}
\item{A decrease of $R_o$ with $p_t$ is expected from initial
  conditions; collective flow only makes this decrease slightly
  stronger.}
\item{The ratio $R_o/R_s$ increases very slowly when one approaches the hydrodynamical limit.
  For the values of $K$ found in Ref.~\cite{Drescher:2007cd},
  it is lower than $1.2$, and much lower than
  predicted by hydrodynamics. Partial thermalization solves most of
  the HBT puzzle.}
\item{For non-central collisions, the variations of  $R_o^2$, $R_s^2$
  and $R_{os}$ with azimuth have almost equal amplitudes. The
  final eccentricity seen in the side radius $R_s^2$ is very close to the
  initial eccentricity. }
\end{itemize}

Our results are in quantitative agreement with data for $R_o/R_s$,
$\Delta R_{os}/\Delta R_o^2$ and $\Delta R_o^2/\Delta R_s^2$. 
On the other hand, our absolute values for $R_o$ and $R_s$ are much
too small.  
Experimentally, it is also found that the final eccentricity is
smaller than the initial eccentricity, almost by a factor 2. 
Both effects cannot be due to flow alone. 
On the other hand, they might be a signature of the softness of the
QCD equation of state or, equivalently, of the transition from a
quark-gluon plasma to a hadron gas. When the quark-gluon plasma
transforms into hadrons, the volume of the system increases by a large
factor: the source swells, which results in larger radii and a smaller
eccentricity.

\section*{Acknowledgments}

C. G. and J.Y.O. thank M. Lopez Noriega, M. A. Lisa, S. Pratt and Yu. 
Sinyukov for useful discussions. 
T.L. thanks W. Florkowski for discussions.
T.L. is supported by the Academy of Finland, contract 126604. 

\bibliographystyle{h-physrev4mod2}
\bibliography{spires}

\end{document}